\documentclass[12pt]{iopart}

\usepackage{graphicx}
\usepackage[breaklinks=true,colorlinks=true,linkcolor=blue,urlcolor=blue,citecolor=blue]{hyperref}

\begin{document}


\title[The role of the hadron initiated single electromagnetic subcascades in IACTs]{The role of the hadron initiated single electromagnetic subcascades in IACT observations}

\author{Dorota Sobczy\'{n}ska}

\address{University of {\L }\'{o}d\'{z}, 
Department of Astrophysics, Pomorska 149/153, 90-236 {\L }\'{o}d\'{z}, Poland}
\ead{dsobczynska@uni.lodz.pl}

\begin{abstract}
The sensitivity of Imaging Air Cherenkov Telescopes (IACTs) worsens significantly at low energies because the $\gamma$/hadron separation becomes much more complex. In this paper we study the impact of the single electromagnetic subcascade events on the efficiency of the $\gamma$/hadron separation for a system of four IACTs using Monte Carlo simulations. The studies are done for: two different altitudes of the observatory, three different telescope sizes and two hadron interaction models (GHEISHA and FLUKA). More than 90$\%$ of the single electromagnetic proton-induced subcascade events are showers with primary energy below 200~GeV, regardless on the trigger threshold. The estimated efficiency of the $\gamma$/hadron separation using the FLUKA model is similar to results obtained using the GHEISHA model. Nevertheless, for at least one triggered telescope only, a higher fraction of single electromagnetic subcascade events was obtained from the FLUKA model. Finally, the calculated quality factors are anti-correlated with the contributions of the false $\gamma$-ray events in the proton initiated showers. Therefore, the occurrence of single electromagnetic subcascade events is one of the main reasons of the worsening of the primary $\gamma$-ray selection efficiency at low energies.

\end{abstract}



\submitto{\JPG}
\maketitle

\section{Introduction}
The discovery of the first~TeV $\gamma$-ray source (the Crab Nebula) in 1989 by the Whipple collaboration \cite{whipple} initiated a fast development of ground-based $\gamma$-ray astronomy. The Imaging Air Cherenkov technique has been successfully used since the first source was discovered. Imaging Air Cherenkov Telescope (IACT) measure the Cherenkov light from Extensive Air Showers (EAS). The Cherenkov photons that are reflected by the telescope mirror are recorded by a matrix of photomultipliers (the so-called telescope camera) mounted in the focal plane of IACT. The shower image that is formed in the camera is a two dimensional angular distribution of the Cherenkov light. 
The number of the registered hadron induced events (the so-called background) is several orders of magnitude larger than the number of the registered $\gamma$-rays events from a source. In 1985 Hillas proposed a method to select  $\gamma$ rays out of a hadron dominated event sample \cite{hillas}. This method is based on parametrization of the recorded shower image. 
The arrival direction of the primary particle is determined by the direction of the image main axis for the data taken with a single IACT and by the intersection of the major axes of both recorded images for a stereo observation. Main axes of the primary $\gamma$-ray images from point-like source are directed to the source position on the camera, while the hadronic background is isotropically distributed. The $\gamma$/hadron separation methods which are used now are more sophisticated (such as \cite{kraw2006,random2008,ohm09,par14}), but most of them are still based upon the original Hillas parameters.

Currently operating observatories, such as H.E.S.S. \cite{aha06}, MAGIC \cite{perf14a, perf14b} and VERITAS \cite{weekes2002,hold11} use arrays of IACTs with large mirror areas. The potential sources of the $\gamma$ rays are observed in stereo mode, in order to improve the sensitivity of telescopes. The Cherenkov Telescope Array (CTA) Collaboration {\cite {actis11,acha13}} plans to build arrays of telescopes with different sizes to measure the fluxes of $\gamma$-ray sources in a large energy range from a few tens of~GeV to hundreds of~TeV. The detection of a low energy showers is possible by using very large telescopes, with a parabolic shape that avoid broadening of the time profile of the Cherenkov signal. It has been shown, based on real data (see e.g. \cite{crab2008}), that the $\gamma$/hadron separation in IACTs becomes much more difficult below 100~GeV. This is motivated by a few effects. At first, larger fluctuations of the Cherenkov light density at ground are expected in a low energy region \cite{bhat,sob2009} which results in larger fluctuations of the image parameters. Second, the geomagnetic field influences the image parameters \cite{comm08,szan13}. Thirdly, IACTs can be triggered by a primary electron or positron initiated shower \cite{elhess}. There are no physical reasons for the differences between the images of a pure electromagnetic subcascade induced by a primary $\gamma$ ray and by an electron. Therefore this kind of background cannot be rejected by using the Hillas parameters describing the image shape.
Fourthly, it has been suggested in \cite{maier} and shown in \cite{sob2007,sob2009b,sob2010} that a specific type of hardly reducible background occurs when observing the low energy region: telescopes can be triggered by light produced by electrons and positrons from only one or two electromagnetic subcascades, which are products of a single $\pi^0$ decay in the hadron initiated shower. 
Shower images formed by Cherenkov photons from a single electromagnetic subcascade can be called false $\gamma$-ray events because they have a very similar shape to the primary $\gamma$-ray images. These false $\gamma$-ray images can be slightly narrower because they start deeper in the atmosphere than real $\gamma$ rays. Therefore a narrower angular distribution of secondary $e^+$ and $e^-$ is expected (see e.g. \cite{giller}). The efficiency of the $\gamma$/hadron separation method, based on the parameters describing the image shape, deteriorate at low energy due to the occurrence of the false $\gamma$-ray events. However, the orientation of the major image axis of the false $\gamma$-ray events is randomly distributed. Therefore parameters, that determine the shower direction, are still effective variables for the primary $\gamma$-rays selection.

The results presented in this paper are based on the Monte Carlo (MC) simulations for a system of four IACTs. We investigate the impact of the occurrence of false $\gamma$-ray events on the efficiency of the $\gamma$/hadron separation. The results were obtained for: i) two hadron interaction models (GHEISHA and FLUKA); ii) two different altitudes of the observatory ($2.2\; {\rm km}$ and $4\; {\rm km}$ a.s.l.); iii) three azimuth angles of $0^{\circ}$, $53^{\circ}$ and $82^{\circ}$; iv) three different telescope areas ($230\; {\rm m^{2}}$, $160\; {\rm m^{2}}$ and $100\; {\rm m^{2}}$). 
In the following we present the MC study and show the fraction of the false $\gamma$-ray candidates in the proton initiated showers for different multiplicities of triggered telescopes. The fraction of single electromagnetic subcascade in the total proton background is estimated for energy range larger than used in the simulations. This fraction is calculated for different trigger thresholds. We show how the contribution of single electromagnetic subcascade in the triggered proton events depends on the average of SIZE (this is the sum of all signals from pixels which belongs the image). 
The scaled WIDTH, scaled LENGTH \cite {daum97} and the height of the shower maximum are applied for the selection of $\gamma$ rays out of the hadron induced showers. We calculate the quality factor (QF) in order to demonstrate the $\gamma$-ray selection efficiency. The strong anti-correlation between QF and the contribution of the false $\gamma$ rays was found for all simulated sets of the parameters of the IACT system. Therefore, the occurrence of the false $\gamma$-ray events is an important reason of the reduction of the $\gamma$/hadron separation efficiency in the range of small average SIZEs. Finally in the conclusions, we compare the results which were obtained from different interaction models, altitudes of the observatory and telescope sizes. 

\section{Description of the Monte Carlo simulations}

\begin{figure}
\begin{center}
\includegraphics*[width=7cm]{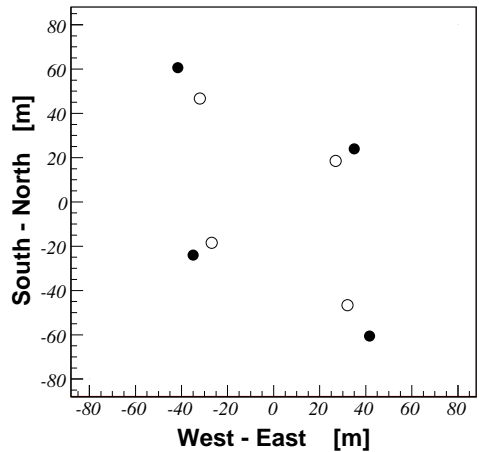}
\end{center}
\caption{The layout of the telescope system on the ground. Filled and open circles correspond to IACT positions at the observation altitude of $2.2\; {\rm km}$ and $4\; {\rm km}$, respectively.}
\label{geom}
\end{figure} 

The investigated system of IACTs contains four telescopes placed in the corners of a diamond having a side length of $85\; {\rm m}$ and diagonals of $85\; {\rm m}$ and $147\; {\rm m}$ and with the altitude of the observation equal to $2.2\; {\rm km}$ a.s.l. (see filled circles in Figure 1). Both the altitude of the observatory and the distance between the two closest telescopes has been chosen as they are for the actual MAGIC telescopes configuration \cite{ale12}. The strength and direction of the geomagnetic field is fixed to that at MAGIC site (La Palma). The side length of the diamond and the distance between the two closest telescopes are reduced to $64\; {\rm m}$ for the simulated altitude equal to $4\; {\rm km}$ a.s.l. (see open circles in Fig.~1). Each simulated telescope is very similar to the MAGIC telescope \cite{ale12}. It has a parabolic shape with a focal length of $17\; {\rm m}$. The mirror dish has a diameter of $17\; {\rm m}$. The total reflector area covers approximately $230\; {\rm m^{2}}$. The camera has hexagonal shape that covers in total almost $4^{\circ}$. It consists of 1141 photomultipliers (PMT) with the diameter of $0.1^{\circ}$. 631~pixels (in the inner part of the camera) are used for the trigger. The trigger requires a time coincidence and neighbour logic. All four telescopes from the system are exactly the same. The simulations of telescopes smaller than $230\; {\rm m^{2}}$ were obtained by reducing only the mirror area of the described above IACT. In reality, smaller telescopes are designed with smaller focal length as the Davies-Cotton design \cite{davcot} (e.g. $107\; {\rm m^{2}}$ IACTs in H.E.S.S.) to avoid the so-called coma effect, which is observed for a telescope with the parabolic shape. In cases of the Davies-Cotton design one may expect higher trigger rate and better Hillas parameter determinations for both hadron and $\gamma$-ray events. However, the assumption of smaller telescope area adopted in this paper leads to reducing of the coma effect. Therefore, it is good approximation to demonstrate the impact of telescope size on single electromagnetic subcascade effects.

The shower development in the atmosphere was simulated using the CORSIKA code \cite{heck}. Two CORSIKA versions 6.023 and 6.99 have been used. In the version 6.023 the GHEISHA \cite{gheisha} and VENUS \cite{venus} interaction models have been applied for the low (i.e. for particles with primary momentum below 80~GeV/c) and high energy ranges. In the version 6.99 FLUKA \cite{fluka} and QGSJET-II {\cite{ostap06a,ostap06b}} have been used as the low and high energy interaction models, respectively. 
The standard CORSIKA code has been adapted in order to study the single subcascade events in the proton initiated shower. The information about each subcascade that has been produced in the EAS was kept in modified code. Additionally, the type of the charged particle, which was responsible for the creation of Cherenkov photons, was also saved in the output of the program.

Showers initiated by primary protons were simulated in the energy range between 30~GeV and 1~TeV with a differential spectral index of -2.75 (which is the index of the primary proton spectrum \cite{bess,hareyama}). The impact parameter was distributed randomly within a circle of 1.2~km radius (measured in the plane perpendicular to the shower axis) around the geometrical center of the telescopes system. The proton induced showers were simulated within a cone with a half-opening angle of $5.5^{\circ}$ at a zenith angle of $20^{\circ}$.

The impact parameter was randomly distributed within a circle of $350\; {\rm m}$ radius for the simulations of showers induced by the primary $\gamma$ rays that have an energy range between 10~GeV and 1~TeV . The differential spectral index was chosen to be equal to -2.6 (which corresponds to the index of the Crab spectrum \cite{crab2008}). The direction of the $\gamma$-ray showers was fixed to a zenith angle equal to $20^{\circ}$ that is also the direction of the axes of all telescopes.
 
Seven sets of simulations were performed: (I) one using FLUKA and QGSJET-II as hadron interaction models (code version from 2011); (II, III and IV) three for different azimuth angles ($0^{\circ}$ - showers directed to the north, $53^{\circ}$, $82^{\circ}$) using the GHEISHA and VENUS as the hadron interaction models ; (V) one for the altitude of the observatory equal to $4\; {\rm km}$ a.s.l.; (VI and VII) two for smaller telescope sizes ($160\; {\rm m^{2}}$ and $100\; {\rm m^{2}}$). Different azimuth angles were simulated in order to show how the geomagnetic field influences the occurrence of false $\gamma$-ray events. It has been shown in \cite{szan13} that the effects of geomagnetic field are scaled by the component of B that is perpendicular to the shower axis ($B_{\perp}$). The azimuth angle of $0^{\circ}$, $53^{\circ}$ and $82^{\circ}$ correspond to $B_{\perp}$ equal to 20, 25 and 30 mT, respectively. The distances between the closest telescopes in the plane perpendicular to the shower axis are similar for all simulated azimuth angles and they are in the range between $80\; {\rm m}$ and $85\; {\rm m}$. Overview of parameters and the numbers of simulated events in all MC sets are presented in the Table 1. Note, that showers initiated by primary $\gamma$ rays are simulated in the same way by the two CORSIKA versions considered.
\begin{center}
\begin{table}
\caption {\label{tab1} Overview of parameters and the number of simulated events in all MC sets.}
\begin{indented}
\item[]\begin{tabular}{@{}*{6}{llllll}}
\br
MC&CORSIKA&Azimuth&altitude&telescope&number&number\\
set&version&angle&of observ.&size&of protons&of $\gamma$ rays\\
&& (deg.)&(km a.s.l.) &(m$^{2}$) &(in 10$^{6}$)&(in 10$^{6}$)\\
\mr
I&{\bf 6.990}&0&2.2& 230& 22&0\\
II&{\bf 6.023}&{\bf 0}&{\bf 2.2}& {\bf 230}& 22&1\\
III&6.023&{\bf 53}&2.2& 230& 22&1\\
IV&6.023&{\bf 82}&2.2& 230& 22&1\\
V&6.023&0&{\bf 4.0}& 230& 22&1\\
VI&6.023&0&2.2& {\bf 160}& 22&1\\
VII&6.023&0&2.2& {\bf 100}& 22&1\\
\br
\end{tabular}
\end{indented}
\end{table} 
\end{center}

The detector simulation consists of two parts: The first part includes the full geometry of the mirrors and their imperfections. Additionally in this program, Rayleigh and Mie scattering of Cherenkov light in the atmosphere were taken into account according to the Sokolsky formula \cite{sokol}. The camera properties, such as additional reflections in Winston cones, photocathode quantum efficiency \cite{bario} and its fluctuation, and the photoelectron collection efficiency were considered in the second part of detector simulations.

The level of the night sky background (NSB) has been chosen to that measured at La Palma \cite{nsb}. The NSB was included in the simulation for the trigger conditions only. In reality photoelectrons (p.e.) produced by light of the NSB are randomly distributed over all pixels of the camera. Due to that, all pixels which contain only the signals from NSB should be subtracted before calculating the parameters of the shower image. The commonly used cleaning procedure neglects pixels with signals smaller than chosen values for the core and boundary pixels, which belong to the image. The efficiency of the separation between real and false $\gamma$-ray events may depend on both NSB and cleaning levels. The results obtained without the NSB light and cleaning are more general because they do not depend on observation conditions and the particular cleaning procedure.
A single telescope was triggered by a shower if the output signals in three next neighbouring pixels (3~NN) exceed a certain threshold within a time window of 3 ns. 3~NN trigger logic was applied in all simulations presented in this paper. In reality, different trigger multiplicity might be used for observations with a large single IACT. The simulated pulse of a single photoelectron has a full width at half maximum equal to 3 ns. Three trigger thresholds were studied in this paper: 3, 4, 5 photoelectrons, which correspond to the signal of 3, 4 and 5~p.e. arriving exactly at the same time, respectively. Finally the possible coincidences between triggers of 2, 3 or 4 telescopes were also checked. 
\section{Results and discussion}
\subsection{Fraction of the false $\gamma$-ray events in the total proton background}
In the results presented below we call false $\gamma$-ray events (or images) that fulfilled conditions: i) the light contribution from the hadronic and muonic part of the shower is less than 10$\%$ of the image SIZE, ii) the rest of the image SIZE is the light contribution from a single electromagnetic subcascade in the proton induced shower, regardless of the number of triggered telescopes. According to this definition, a single photoelectron from other electromagnetic subcascades cannot give the contribution to recorded images. Therefore results presented below show the smallest possible impact of single electromagnetic events on observations with IACTs. In particular, a total fraction of false $\gamma$ rays estimated in this paper should be treated as a lower limit.

The distributions of the number of electromagnetic subcascades giving a contribution to all registered images are presented in Figure 2. All distributions are normalized to~1. The height of the second histogram bin corresponds to the fraction of the false $\gamma$-ray candidates only, because conditions i) and ii) are not required. Figure 2a shows plots obtained from MC set II (using the GHEISHA model) for the trigger threshold equal to 3~p.e.. The fraction of the false $\gamma$-ray candidates decreases with the multiplicity of the triggered IACTs as expected. We have checked that the presented distributions are similar for all simulated azimuth angles: $0^{\circ}$, $53^{\circ}$ and $82^{\circ}$ (MC sets II, III and IV). The comparison between the results of the GHEISHA (black) and FLUKA models (grey) are demonstrated in Fig.~2b for the multiplicity of triggered telescopes equal to 1 (long dashed) and 2 (solid histograms) and the trigger threshold of 3~p.e.. The FLUKA model gives higher fraction of the false $\gamma$-ray candidates than the GHEISHA model only for the required number of triggered telescopes equal to 1. The results of both interaction models are very similar when more than one IACT is triggered. The higher fraction of the false $\gamma$-ray candidates was obtained for the simulated altitude equal to $4\; {\rm km}$ a.s.l. (grey histograms in Fig.~2c) than for $2.2\; {\rm km}$ a.s.l. (black histograms), regardless of the required number of triggered telescopes. The proton induced showers contain slightly smaller fraction of the false $\gamma$-ray candidates for a smaller mirror area (e.g. $100\; {\rm m^{2}}$ in Fig.~2d plotted as grey) than for $230\; {\rm m^{2}}$ telescopes (shown as black histograms in this Figure). Significantly higher fraction of events without any electromagnetic contribution (a height of the first histogram bin in Fig.~2) is expected for smaller than for larger telescopes when less than 4 IACTs triggered.

\begin{figure}
\begin{center}
\includegraphics*[width=14cm]{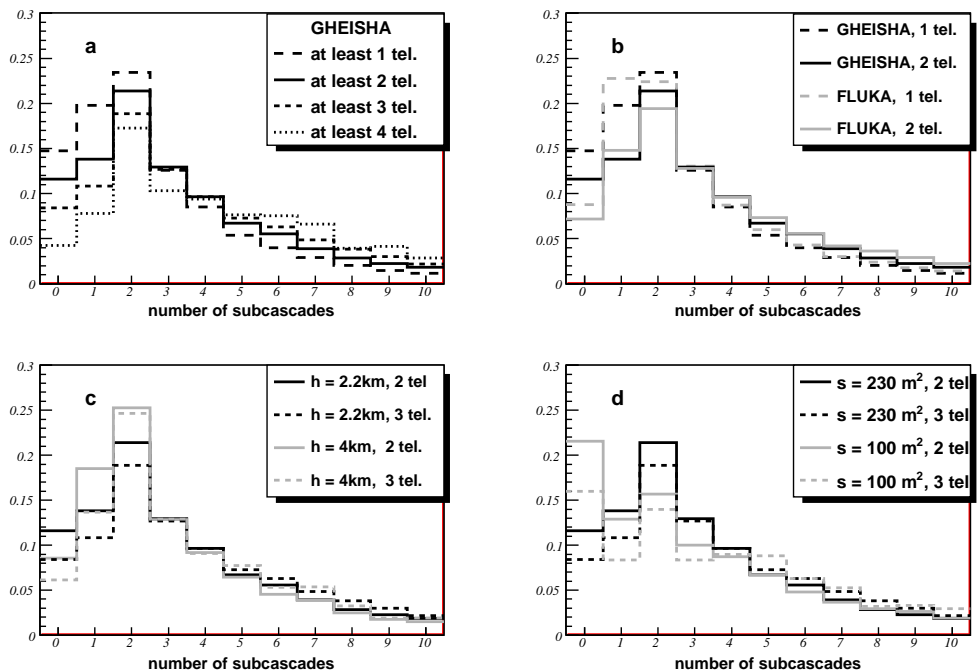}
\end{center}
\caption{Distribution of the number of the electromagnetic subcascades giving a contribution to all registered images of the proton initiated showers for the trigger threshold of 3~p.e.. {\bf a)} The GHEISHA model for all investigated multiplicities of triggered IACTs. {\bf b)} Comparison between interactions models (GHEISHA and FLUKA). {\bf c)} Comparison between two observation altitudes: 2.2 and 4~km a.s.l.. {\bf d)}~Comparison between two telescope sizes: $230\; {\rm m^{2}}$ and $100\; {\rm m^{2}}$. Long dashed, solid, dashed and dotted histograms correspond to the required multiplicities of triggered telescopes from 1 to 4, respectively. All presented distributions were normalized to~1. }
\label{number_subcas}
\end{figure}

We have checked that the expected trigger rate of proton showers is higher for FLUKA in comparison to GHEISHA model when only at least one or two telescopes fulfilled the trigger conditions. For larger required multiplicity of triggered IACTs, the results of both investigated interaction models do not differ significantly. At the altitude of $4\; {\rm km}$ a.s.l. more proton induced events triggered the IACT system than at $2.2\; {\rm km}$, as expected. The trigger rate is lower for smaller mirror area than for $230\; {\rm m^{2}}$ IACT, as expected. 
We have found that the energy distributions of events that fulfilled both i) and ii) conditions are much steeper than that of all triggered showers for all simulation sets. Most of the false $\gamma$-ray events are the proton showers with low energy. For less than four triggered IACTs, more than 94$\%$ of the false $\gamma$-ray images with the SIZE parameter larger than 20~p.e. are the proton induced showers with primary energy below 200~GeV, regardless on the simulation set. This fraction is slightly below 90$\%$ only in case of all triggered telescopes, where obtained trigger energy thresholds for primary proton (the position of the maximum of the energy distribution) are the highest. Note, that the discussed fraction almost does not depend on the chosen interaction model (FLUKA or GHEISHA), magnetic field, observation altitude and mirror area. 

Figure 3 shows how the ratio of the number of false $\gamma$-ray events to the number of all triggered proton showers depends on the energy of the primary proton for the trigger threshold of 3~p.e.. This ratio was calculated separately in each histogram bin. The results obtained from MC set II are presented in Fig.~3a for required multiplicities of triggered telescopes from 1 to 4. The fraction of the false $\gamma$ ray in the energy bin decreases with the multiplicity of triggered IACTs. This fraction also diminishes with energy for all simulation sets (see comparisons between: interaction models in Fig.~3b, different altitudes of the observatory in Fig.~3c and different telescope areas in Fig.~3d). The first bin of presented histograms corresponds to the energy range between 30~GeV and 40~GeV. At these energies one should expect that even up to 45$\%$ of protonic background is single subcascade events at the altitude of $4\; {\rm km}$ a.s.l. for at least two triggered telescopes (see grey solid histogram in Fig.~3c). In the energy range between 250~GeV and 400~GeV, no more than 2$\%$ of all triggered protonic background are false $\gamma$-ray events for the required number of triggered telescopes larger than 1, regardless on the simulation set (see the sixth bin of histograms presented in Fig.~3).

\begin{figure}
\begin{center}
\includegraphics*[width=16cm]{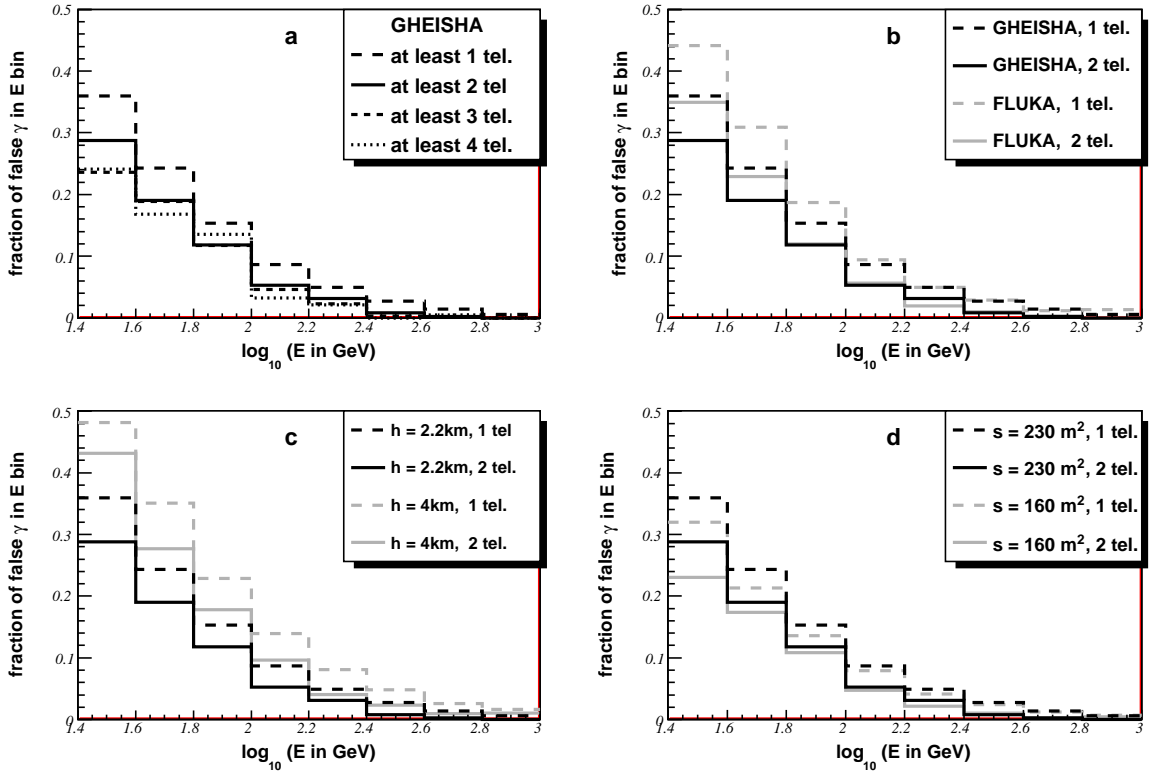}
\end{center}
\caption{The ratio of the number of false $\gamma$-ray events to the number of all triggered proton shower versus primary energy for the trigger threshold equal to 3~p.e.. {\bf a)} The GHEISHA model for all investigated multiplicities of triggered IACTs. {\bf b)} Comparison between interactions models. {\bf c)} Comparison between two observation altitudes: 2.2 and 4~km a.s.l.. {\bf d)} Comparison between two telescope sizes: $230\; {\rm m^{2}}$ and $160\; {\rm m^{2}}$. Long dashed, solid, dashed and dotted histograms correspond to the required multiplicities of triggered telescopes from 1 to 4, respectively. The ratio was calculated in each histogram bin separately.}
\label{energy}
\end{figure} 

To estimate a fraction of false $\gamma$-ray events in the total proton background, one may evaluate the expected number of the triggered and false $\gamma$-ray events for energies above 1~TeV. A simple power-law fits of the energy distribution tail (for images with the SIZE parameter larger than 20~p.e.) was used as an extrapolation function. Energies above 500~GeV and 200~GeV were considered as tails of the energy distribution for all triggered and false $\gamma$-ray events. 

 We have checked that the fraction of false $\gamma$-ray events does not depend on the distance between the two triggered telescopes. Similar values were obtained for the shortest and the longest distance between IACTs. Figure 4 shows how the fraction of the false $\gamma$-ray images in the proton induced showers decreases with the trigger threshold. Four panels in this Figure present comparisons between the results obtained from different MC sets. The fractions of the false $\gamma$-ray events for azimuth angles $0^{\circ}$ (black) and $82^{\circ}$ (grey lines) are plotted in Fig.~4a. The absolute differences are smaller than 1$\%$. Therefore we conclude that estimated fractions do not depend on the simulated azimuth angle. In the investigated range of $B_{\perp}$ (between 20 mT and 30 mT) the geomagnetic field does not influence any of the two numbers: all triggered proton and false $\gamma$-ray events. This conclusion is consistent with the results shown in \cite{szan13}, where authors claimed that geomagnetic field has much smaller impact on the proton than on $\gamma$-ray initiated showers even for the Large Size Telescope (LST) which is proposed in the CTA project \cite{actis11}. 

A comparison of the interaction models is presented in Figure 4b. In case of the observations with a single telescope (long dashed lines), the contribution of false $\gamma$-ray images in all triggered events decreases from approximately 11$\%$ to 7$\%$ when the trigger threshold increases from 3~p.e. to 5~p.e. for the FLUKA model (grey) and from 9$\%$ to 5$\%$ for the GHEISHA model (black lines). For at least 2 triggered IACT, the estimated fraction is still higher for FLUKA than for GHEISHA, however absolute differences are smaller than 1$\%$. For larger multiplicities the differences between used interaction models are negligible. 
It has been shown in \cite{maier} that the probability that the majority of the primary energy is deposited in the electromagnetic part of the shower is lower for the GHEISHA than for FLUKA model. Due to that, the estimated fraction of the false $\gamma$ rays in the protonic background is lower for GHEISHA than for FLUKA model for at least one triggered telescope. However, if more IACTs are triggered then similar results were obtained for both interaction models. This may be explained by the fact that some Cherenkov photons from other electromagnetic subcascades can contribute to registered images. The probability of this effect is higher for FLUKA which produces larger electromagnetic part of the shower, while, the probability that more telescopes fulfilled trigger conditions decreases in a similar way for both interaction models.

The estimated fraction of false $\gamma$-ray events for the altitude equal to $4\; {\rm km}$ (grey lines in Fig.~4c) are higher than for $2.2\; {\rm km}$ (black lines in the same Figure) in the investigated ranges of the trigger threshold and telescope multiplicity. Significantly more false $\gamma$-ray events are observed at higher altitudes, regardless on the multiplicity of triggered telescope. Therefore, the expected fraction of the false $\gamma$-ray events in the proton initiated showers is strongly influenced by the altitude of the observatory.

Figure~4d shows a comparison between the contributions of false $\gamma$-ray events estimated for two simulated mirror areas: $230\; {\rm m^{2}}$ (black) and $100\; {\rm m^{2}}$ (grey lines). At fixed trigger threshold, this fraction significantly increases with the mirror area, although the estimated fraction of the false $\gamma$-ray candidates only slightly increases with the telescope size (see Fig.~2d). However for smaller mirror areas, large fraction of those candidates contains photoelectrons created by Cherenkov photons from hadronic and muonic parts of the shower. In cases of telescope multiplicities lower than 3, the fraction of the false $\gamma$-ray events obtained for $230\; {\rm m^{2}}$ at trigger threshold of 5~p.e. are very close to results for $160\; {\rm m^{2}}$ telescope at trigger threshold of 4~p.e. and for $100\; {\rm m^{2}}$ telescope at trigger threshold of 3~p.e.. The simulations of the IACT with $100\; {\rm m^{2}}$ mirror show that the expected fraction of the false $\gamma$-ray events is negligible for the multiplicity of triggered telescope larger than 2 (see grey dashed and dotted lines in Fig.~4d). In those cases energy thresholds obtained for proton showers are above 250~GeV. 
In this energy range the probability of triggering the IACT system by a single electromagnetic subcascade is very small even for at least one triggered telescope.

\begin{figure}
\begin{center}
\includegraphics*[width=16cm]{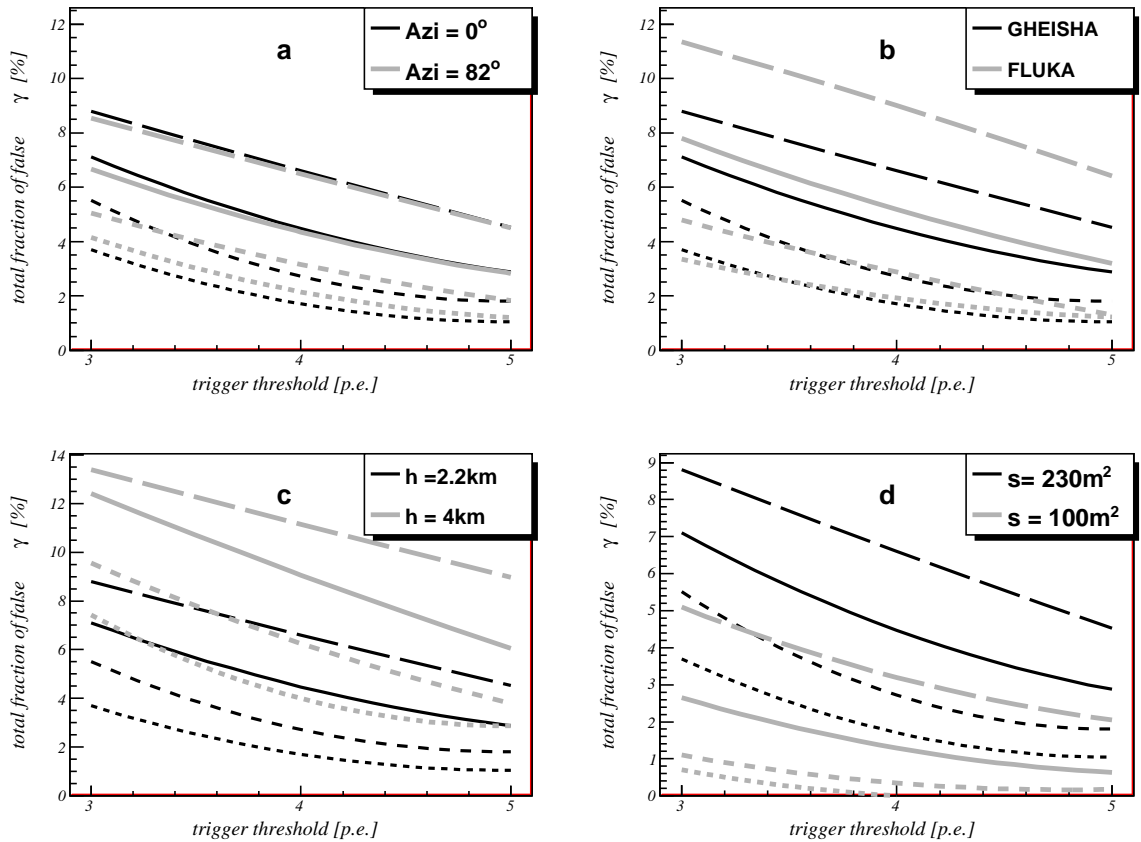}
\end{center}
\caption{Fraction of the false $\gamma$-ray events in the total protonic background as a function of the trigger threshold for: {\bf a)} two different azimuth angles (from MC simulation sets II and IV); {\bf b)} GHEISHA/FLUKA model (from MC sets II and I); {\bf c)}~two different altitudes of the observatory: $2.2\; {\rm km}$ a.s.l. and $4\; {\rm km}$ a.s.l. (from MC sets II and V); {\bf d)} two different telescope areas (from MC sets II and VII). In all panels, long dashed, solid, dashed and dotted lines correspond to the multiplicity of the triggered IACTs equal to 1, 2, 3 and 4, respectively. MC sets are defined in Table 1.}
\label{tot_frac}
\end{figure} 

\subsection{Efficiency of the $\gamma$-rays selection}

The SIZE parameter was calculated for each triggered telescope. An average SIZE ($<$SIZE$>$) of the event was obtained from parameters measured by each triggered IACT. We have checked that more than 90$\%$ of false $\gamma$-ray events have $<$SIZE$>$ below 300~p.e. for the multiplicity of triggered telescopes lower than 3. Figure 5 shows how the ratio of the number of the false $\gamma$-ray events to the number of all triggered proton showers depends on the $<$SIZE$>$. This ratio was calculated separately in each histogram bin for the trigger threshold of 3~p.e.. The results of MC set II are shown in Fig.~5a. The fraction of the false $\gamma$-ray events in $<$SIZE$>$ bin does not depend on the multiplicity of triggered IACTs (see long dashed, solid and dashed histograms in this Figure). In the lowest presented $<$SIZE$>$ bin approximately 30$\%$-40$\%$ of the proton initiated showers are the false $\gamma$-ray events. This fraction decreases with $<$SIZE$>$. The results obtained from other MC sets are also very similar for different multiplicities of triggered telescopes. Therefore the comparison between models (see Fig.~5b), altitudes (see Fig.~5c) and telescope sizes (see Fig.~5d) is shown only for at least two triggered IACTs. The histograms obtained with GHEISHA (black) and FLUKA (grey) do not differ significantly. We have checked that in the case of at least one triggered telescope only, the calculated ratio is slightly higher for FLUKA than for GHEISHA model in whole range of the average SIZE. 
It is seen in Fig.~5c that the contribution of the false $\gamma$-ray events in the proton background is higher for the simulated altitude of $4\; {\rm km}$ (grey) than for $2.2\; {\rm km}$ (black histogram). 
The contribution of false $\gamma$-ray events in each chosen $<$SIZE$>$ bin is lower in case of a smaller telescope area (in Fig.~5d grey histogram corresponds to $100\; {\rm m^{2}}$ telescopes). 

\begin{figure}
\begin{center}
\includegraphics*[width=16cm]{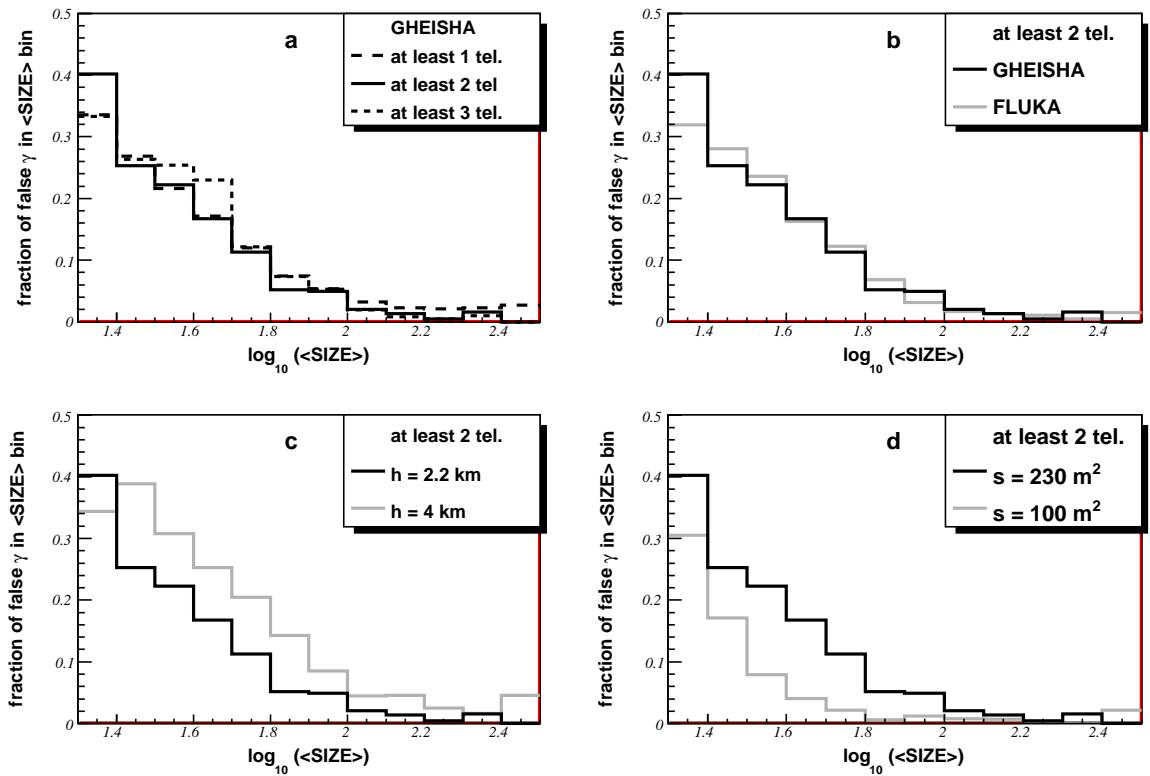}
\end{center}
\caption{The ratio of the number of false $\gamma$-ray events to the number of all triggered proton shower versus $<$SIZE$>$ for the trigger threshold equal to 3~p.e.. {\bf a)}~The GHEISHA model for different multiplicities of triggered IACTs. {\bf b)} Comparison between interactions models for at least 2 triggered IACTs. {\bf c)} Comparison between two observation altitudes: 2.2 and 4~km a.s.l.. {\bf d)} Comparison between two telescope sizes: $230\; {\rm m^{2}}$ and $100\; {\rm m^{2}}$. Long dashed, solid and dashed histograms correspond to the required multiplicities of triggered telescopes from 1 to 3, respectively. The ratio was calculated in each histogram bin separately.}
\label{fr_size}
\end{figure} 

Two Hillas parameters relevant for this study are LENGTH and WIDTH. They are defined as the standard deviations of the image profiles along the long and short axis of the image, respectively \cite{hillas}.
Both parameters can be used for the selection of the $\gamma$-ray events from the hadron initiated showers. The more efficient $\gamma$/hadron methods use the so-called scaled WIDTH (WIDTH$_{S}$) and scaled LENGTH (LENGTH$_{S}$) (for definition see \cite{daum97}). 
The scaling factors are calculated from simulated $\gamma$-ray images only. They are applied to the calculations of the WIDTH$_{S}$ and LENGTH$_{S}$ for the images of $\gamma$-ray and proton initiated showers. As a result, the WIDTH$_{S}$ and LENGTH$_{S}$ distributions of the $\gamma$-ray initiated showers are Gaussian distributions with mean equal to 0 and dispersion equal to 1. For each shower which triggered the system, the mean WIDTH$_{S}$ ($<$WIDTH$_{S}$$>$) and mean LENGTH$_{S}$ ($<$LENGTH$_{S}$$>$) were calculated from scaled parameters obtained from each recorded image. 
The distributions of the $<$WIDTH$_{S}$$>$ and $<$LENGTH$_{S}$$>$ parameters are presented in Figure 6a and 6b for at least two triggered telescopes at the trigger threshold equal to 3~p.e.. The results of MC set II are plotted as an example. The histograms of all triggered events for primary $\gamma$ ray (dashed black) and proton (solid grey) are normalized to~1. Appropriate factors were used to scale the histograms of the false $\gamma$-ray events. 
The distribution of false $\gamma$-ray images (solid black) is shifted towards lower $<$WIDTH$_{S}$$>$ in comparison to the real $\gamma$-ray events, because the detected false $\gamma$-ray events are subcascades which begin deeper in the atmosphere. 
Therefore, the observed parts of electromagnetic cascade are younger and the angular distributions of electrons and positrons are narrower than in the showers initiated by primary $\gamma$ rays (see e.g. \cite{giller}). It can be seen in Figure 6, that one can expect very low efficiency of the primary $\gamma$-rays selection from false $\gamma$-ray events if the WIDTH$_{S}$ and LENGTH$_{S}$ parameters only are used in the separation method.

\begin{figure*}[t]
\centerline{\includegraphics[width=5cm]{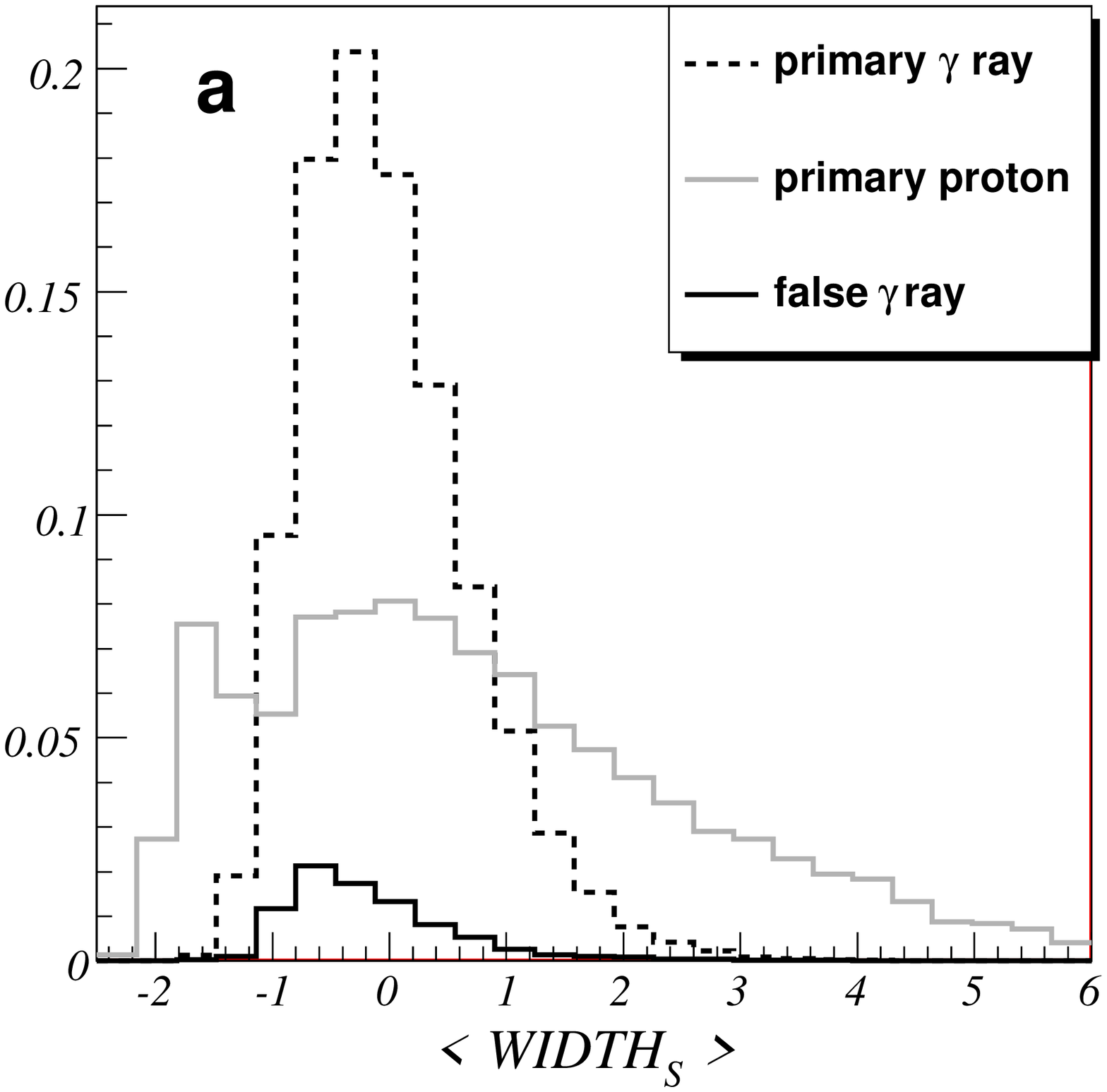}
\includegraphics[width=5cm]{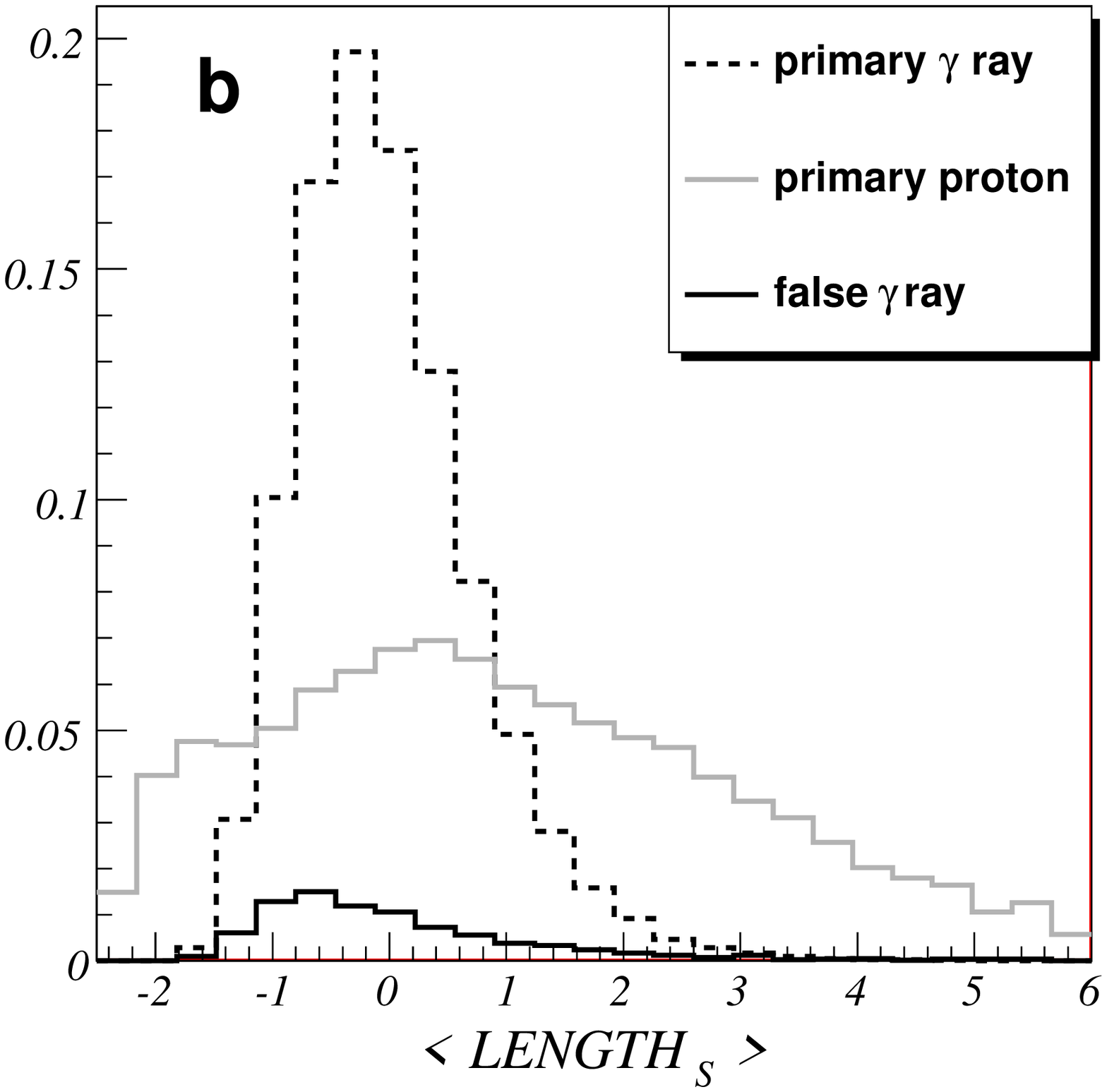}
\includegraphics[width=5cm]{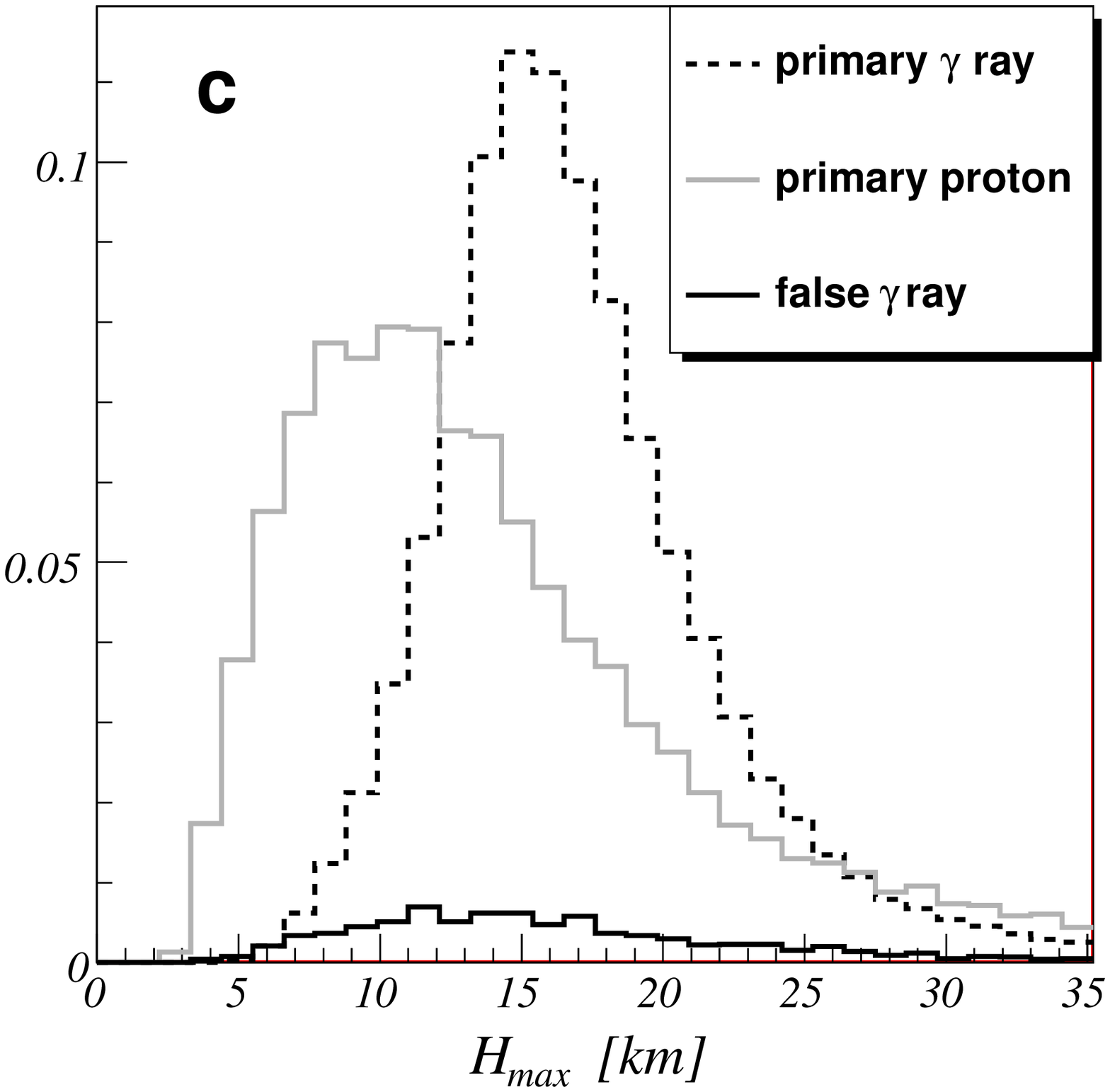}
}
\caption{The distributions of the parameters obtained from MC set II: {\bf a)} the mean WIDTH$_{S}$; {\bf b)} the mean LENGTH$_{S}$; {\bf c)} and the height of the shower maximum. The primary $\gamma$-ray initiated showers (black dashed) and proton initiated showers (grey solid histograms) are shown for observations with at least two IACTs with the trigger threshold equal to 3~p.e.. The histograms for primary $\gamma$ ray and proton are normalized to~1. Appropriate factors were used to scale the histograms of the false $\gamma$-ray events (black solid histograms). Only images with SIZE above 10 p. e. are presented.}
\label{wid_len}
\end{figure*}

The height of the shower maximum (H$_{max}$) can be used to separate $\gamma$ rays from proton events \cite{perf14b}. The distributions of the H$_{max}$ parameter (in km a.s.l.) obtained from MC set II are presented in Figure 6c for at least two triggered IACTs at the trigger threshold of 3~p.e.. The height of the shower maximum is a good parameter to use in the $\gamma$/hadron separation. However, there is no clear separation between real and false $\gamma$-ray results. Therefore, the real and false $\gamma$-ray events cannot be effectively distinguished by using the height of the shower maximum. 

The occurrence of the false $\gamma$-ray events affects the efficiency of the $\gamma$/hadron separation. In order to estimate this effect quantitatively, whole sample of triggered events was split into five intervals of $<$SIZE$>$. Next, the $\gamma$/hadron separation method was applied in each $<$SIZE$>$ bin. 
Primary $\gamma$-ray candidates are selected as events that fulfilled conditions: i) have all recorded images with WIDTH$_{S}$ and LENGTH$_{S}$ in the range between -1.5 and 1.5, ii) have the height of the shower maximum above 9.5~km~a.s.l.. The estimated energy of primary $\gamma$ rays (E$_{est}$) that correspond to each $<$SIZE$>$ interval is given in the Table 2 for different simulation sets. The presented energy was estimated as the peak position of the energy distribution when at least 2 telescopes triggered at trigger threshold of 3~p.e.. 
Note, that for fixed $<$SIZE$>$ bin different values of E$_{est}$ were obtained from different MC sets. Thus the efficiency of the $\gamma$-ray selection presented in this paper was calculated in energy ranges which depend on the simulations set.

\begin{center}
\begin{table}
\caption {\label{tab2} The estimated energy (E$_{est}$) of primary $\gamma$ rays in~GeV.}
\begin{indented}
\item[]\begin{tabular}{@{}*{6}{llllll}}
\br
$<$SIZE$>$  &&(20,50)&(50,100) &(100,200) &(200,300) &$>$ 300\\
 \mr 
 mirr. area & altitude& E$_{est}$& E$_{est}$& E$_{est}$& E$_{est}$ &E$_{est}$\\
 (m$^{2}$)& a.s.l. (km)& (GeV) & (GeV) & (GeV) & (GeV) & (GeV)\\
\mr
230 & 2.2  & 15. & 25. & 50. & 90. & 135.\\
160 & 2.2  & 20. & 35.& 65. & 120. & 195.\\
100 & 2.2  & 30. & 50.& 100. & 185. & 300.\\
230 & 4.0  & 11. & 18.& 30. & 60. & 90.\\
\br
\end{tabular}
\end{indented}
\end{table} 
\end{center}

The quality factor (QF) was used to demonstrate the efficiency of the $\gamma$-ray selection from the hadronic background \cite{aha93}. This factor is related to the significance of the measurement (the ratio of the $\gamma$-ray signal and the dispersion of the Poisson-like hadronic background). The quality factor can be defined as, 
$$Q_{\rm F}\equiv \frac{N^{\rm c}_{\gamma}/{N^{\rm all}_{\gamma}}}{\sqrt{N^{\rm c}_{\rm p}/N^{\rm all}_{\rm p}}},$$
where $N^{\rm all}_{\gamma}$,$N^{\rm all}_{\rm p}$ are the total numbers of triggered events for chosen $<$SIZE$>$ bin and $N^{\rm c}_{\gamma}$, $N^{\rm c}_{\rm p}$ are the numbers of events surviving the $\gamma$/hadron separation cuts. The QF, which were calculated in this paper, take into account only the differences between parameters describing image shapes and differences in H$_{max}$ of the showers initiated by primary $\gamma$ rays and protons. Other parameters that are used in the primary $\gamma$-rays selection are not connected with the occurrence of the false $\gamma$-ray events.

The quality factor versus the $<$SIZE$>$ parameter is shown in Figure 7a for at least two triggered IACT at the triggered threshold of 4~p.e.. The same trigger conditions were chosen to prepare plots in all panels of the Figure 7.
The efficiency of the $\gamma$/hadron separation increases with the $<$SIZE$>$ for all simulation sets as expected. The value of QF decreases with the mirror area - compare the results for $100\; {\rm m^{2}}$ (dashed), $160\; {\rm m^{2}}$ (dotted) and $230\; {\rm m^{2}}$ telescopes (solid line). The quality factors obtained using the FLUKA model (grey solid) are very similar to values calculated using the GHEISHA model (black solid). The efficiency of the primary $\gamma$-ray selection is worse for the altitude equal to $4\; {\rm km}$ (dash-dotted) than for $2.2\; {\rm km}$ (solid line). We have check that the QF obtained for different azimuth angles (i.e. different $B_{\perp}$) do not differ in contrast to the results presented in \cite{comm08,szan13}. This disagreement can be explained by the fact, that in the $\gamma$/hadron separation method we have not use parameters describing the image orientation that were applied by authors in \cite{comm08,szan13}.

The quality factor versus the contribution of false $\gamma$-ray events in protonic background is presented in Figure 7b. The contribution of false $\gamma$-ray events before $\gamma$/hadron separation was calculated in each $<$SIZE$>$ bin separately. The quality factors decreases with the fraction of the false $\gamma$-ray events in a similar way for all simulation sets if the contribution of the false $\gamma$-ray events is larger than~0. 

In order to present how the quality factor is reduced due to occurrence of the background caused by the false $\gamma$-ray events, the ratio between QF and the maximal value of QF (QF$_{max}$) is shown as a function of the contribution of the false $\gamma$-ray events (see Figure 7c). It can be seen, that the results of all MC sets contain the points with negligible contribution of false $\gamma$-ray events (points with coordinates close to (0,1)), except the altitude of the observatory equal to $4\; {\rm km}$ a.s.l.. Therefore for this altitude, QF$_{max}$ is underestimated because it was calculated from the sample of showers with the contribution of the false $\gamma$-ray events approximately equal to 2$\%$. The most rapid reduction of a quality factor with the contribution of the false $\gamma$-ray events is expected for IACT system with the smallest telescopes (diamonds). 
In this case the calculated QF for $<$SIZE$>$ larger than 300~p.e. (i.e. E$_{est}$ equal to 300~GeV) is approximately 5~times higher than for $<$SIZE$>$ between 20 and 50~p.e. (i.e. E$_{est}$ equal to 30~GeV). QF parameter is reduced by a similar factor in case of 160 m$^{2}$ telescopes (stars) and by a factor of 4 for GHEISHA/FLUKA model (filled/open circles) with the corresponding reduction of E$_{est}$ (see numbers in the Table 2). The slowest decline of QF/QF$_{max}$ was obtained for the IACT system located at the altitude equal to $4\; {\rm km}$ a.s.l. (crosses) where the E$_{est}$ is lower than 100~GeV for all $<$SIZE$>$ bins. However, as it was mentioned before, QF$_{max}$ is underestimated for this simulation set.

\begin{figure*}[t]
\centerline{\includegraphics[width=5cm]{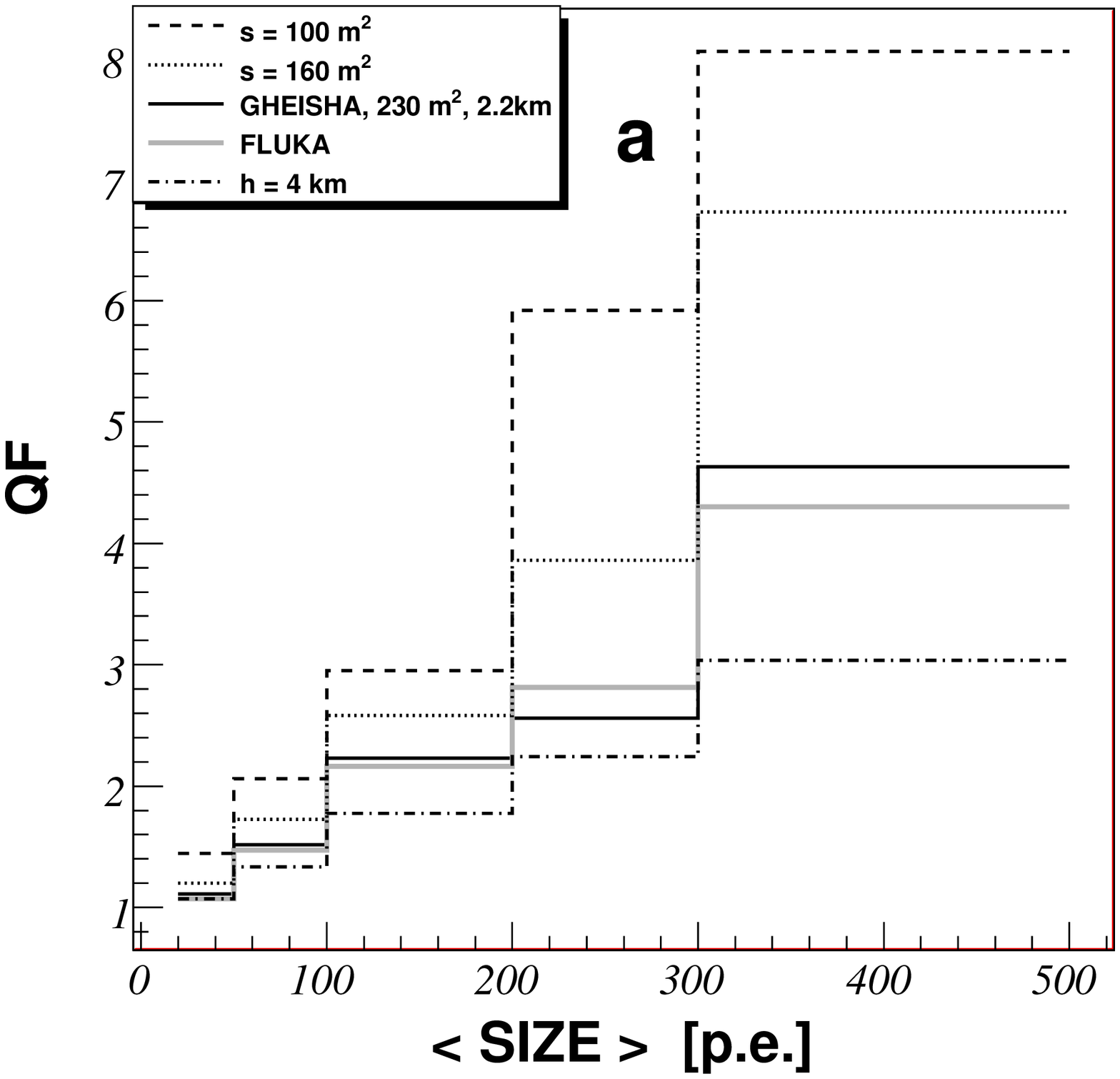}
\includegraphics[width=5cm]{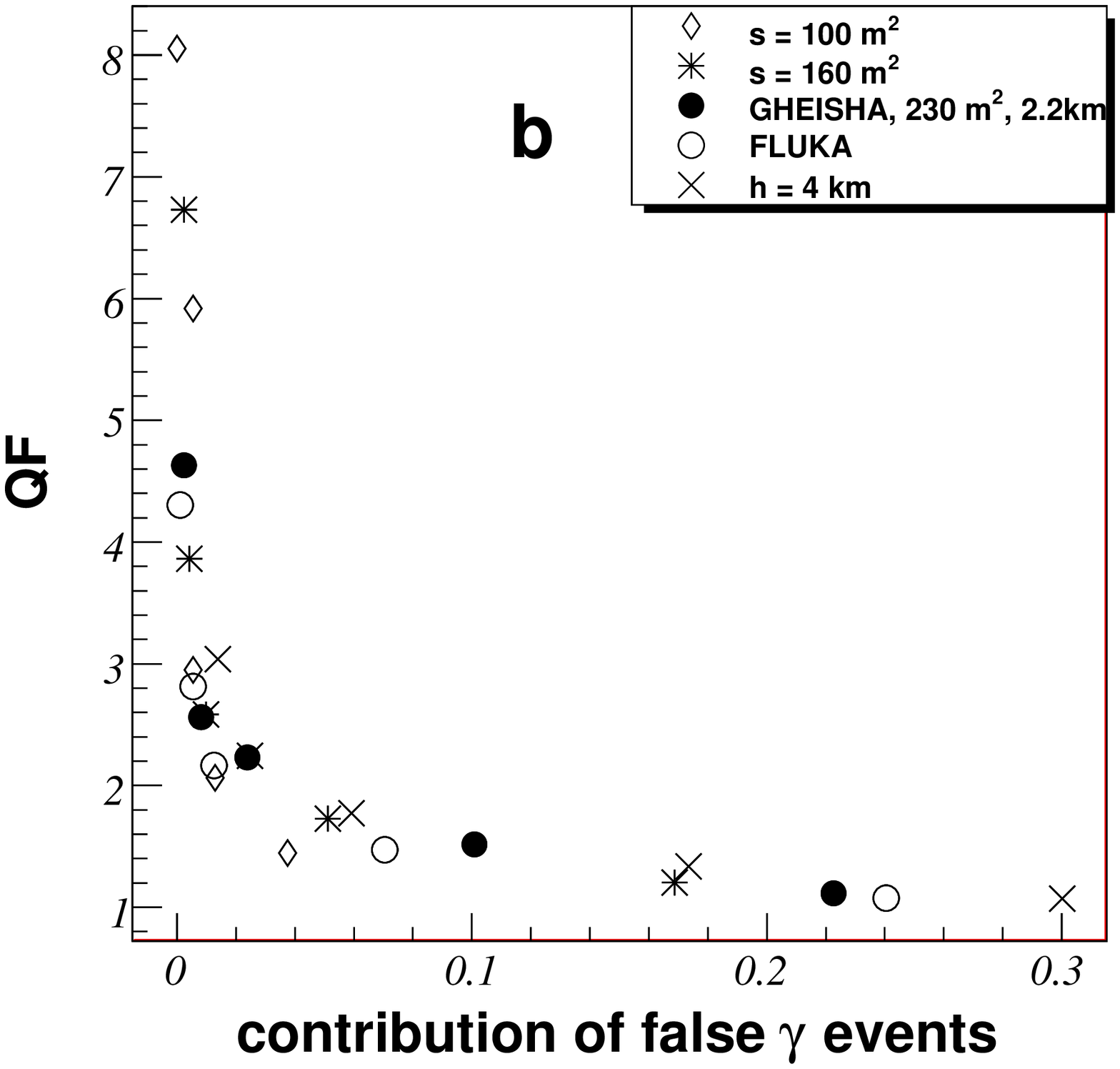}
\includegraphics[width=5cm]{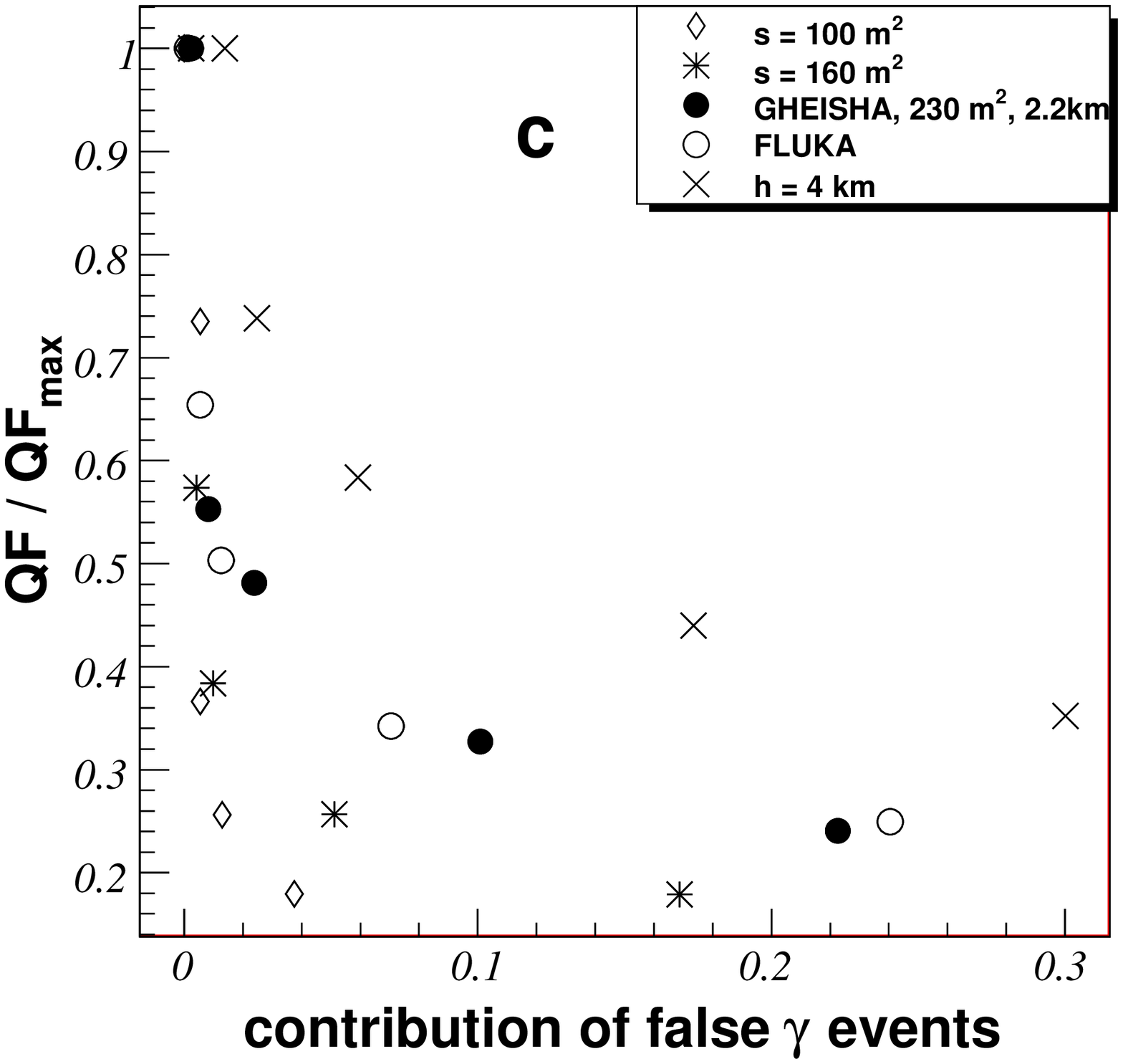}
}
\caption{The quality factor: {\bf a)} versus $<$SIZE$>$; {\bf b)} versus the contribution of the false $\gamma$-ray events in proton background before the $\gamma$/hadron separation. {\bf c)} The ratio between QF and the maximal value of QF versus the contribution of the false $\gamma$-ray events. Diamonds, stars, full/open circles and crosses in Fig.~7b and 7c correspond to simulations of $100\; {\rm m^{2}}$ mirror area, $160\; {\rm m^{2}}$ mirror area, $230\; {\rm m^{2}}$ telescope with GHEISHA/FLUKA model and $4\; {\rm km}$ a.s.l. altitude, respectively. All presented results were obtained for the trigger threshold of 4~p.e. and the required multiplicity of the triggered telescopes equal to 2.}
\label{QF}
\end{figure*}

Finally, we calculated the correlation coefficient between QF and the contribution of false $\gamma$-ray events in protonic background before $\gamma$-ray selection. Correlation coefficient changes significantly with increase of the multiplicity of triggered telescopes from 1 to 4. At the trigger threshold of 4~p.e. it varies from -0.95 to -0.5 and from -0.97 to -0.55 for the GHEISHA and FLUKA model, respectively. For the altitude of the observatory equal to $4\; {\rm km}$, the correlation is stronger as its coefficient changes from to -0.86. In case of 160 m$^{2}$ IACTs, calculated coefficient changed from -0.94 to -0.71 with decreasing the multiplicity of triggered telescopes from 1 to 3. The weakest correlation was obtained for the smallest investigated telescopes. The correlation coefficients are equal to -0.92 and only -0.6 for at least one and 2 triggered telescopes, respectively. In cases of higher multiplicity of triggered IACTs, the correlation coefficient was not calculated because the total fraction of the false $\gamma$-ray events is negligible (see Fig.~4c). 

\section {Conclusions}

The occurrence of false $\gamma$-ray events plays an important role in the observations using an Imaging Air Cherenkov Telescopes system. The study brought two important conclusions. First, this kind of background is mainly caused by proton induced showers with the primary energy below 200~GeV. Second, most of the false $\gamma$-ray events have $<$SIZE$>$ smaller 300~p.e.. Both of those features do not depend on: the interaction model, altitude of the observatory and size of the telescope mirror. Those conclusions are consistent with the results obtained for a smaller trigger area of the camera \cite{sob2009b}. The quality factor depends much stronger on the multiplicity of triggered telescopes than on the trigger threshold because the energy threshold of the IACT system is more sensitive on the required number of triggered telescopes than on the trigger threshold. We have checked, that the effect of the false $\gamma$-ray events does not depend on the geomagnetic field for $B_{\perp}$ in the range between 20 and 30 mT.

The simulations presented in this paper allowed to compare the results obtained with two models of the hadron interactions. We have shown, that for the multiplicity of triggered IACTs larger than 1, there are not significant differences between results obtained using GHEISHA and FLUKA model. 
In case of at least one triggered IACTs, both the estimated total trigger rate and the expected fraction of false $\gamma$-ray events are larger for FLUKA than for GHEISHA model. In this case the contribution of the single electromagnetic subcascade events in each $<$SIZE$>$ intervals is also slightly higher for the FLUKA model. 

Comparison of the results for altitudes of $2.2\; {\rm km}$ and $4\; {\rm km}$ a.s.l. leads to the following conclusions. The system of IACTs located at higher altitude can be triggered by protons at lower energies. Due to that, the fraction of false $\gamma$-ray events is significantly larger for altitude equal to $4\; {\rm km}$ than for $2.2\; {\rm km}$, regardless on the multiplicity of triggered telescopes and trigger threshold. The contribution of single subcascade events in chosen interval of $<$SIZE$>$ is also larger for IACT system located at higher altitude. As a consequence, the efficiency of the primary $\gamma$-ray selection is lower for higher altitude of the observatory independently on the number of triggered telescopes.

The system of IACTs with smaller mirror areas can be triggered by proton showers at higher energies than in the case of a system with larger telescopes. As a result the fraction of single subcascade events also decreases significantly when the mirror area changes from $230\; {\rm m^{2}}$ to $100\; {\rm m^{2}}$. However, for the multiplicity of triggered IACTs lower than 3 the fraction obtained for $230\; {\rm m^{2}}$ at trigger threshold 5~p.e. is comparable with results for $160\; {\rm m^{2}}$ and $100\; {\rm m^{2}}$ at trigger threshold of 4~p.e. and 3~p.e., respectively. The fraction of hardly reducible background for the observations with at least three or at least four triggered telescopes are not simply shifted in the trigger threshold level. The estimated fraction of the false $\gamma$-ray events is negligible (lower than 1$\%$) for the multiplicity of triggered IACTs larger than 2 and the mirror area equal to $100\; {\rm m^{2}}$. Telescopes with such mirror areas are proposed to be applied as the Medium Size Telescopes (MST) in the CTA project. The data taken by at least three MSTs should contain negligible fraction of the background caused by single electromagnetic subcascade events.

We have shown that the efficiency of the $\gamma$/hadron separation is strongly anti-correlated with the contribution of the false $\gamma$-ray events in the proton induced showers. Additionally, the occurrence of the false $\gamma$-ray events has less influence on the data that are taken by at least 3 or 4 triggered telescopes because in this case, the system is triggered mostly by higher energy protons. The results presented in this paper were obtained using a basic reconstruction of the shower. A new, more sophisticated reconstruction algorithm may lead to better $\gamma$/hadron separation. However, the occurrence of the false $\gamma$-ray events is one of the main causes of the deterioration of both the $\gamma$/hadron separation efficiency and as a result sensitivity of a system of IACTs in the low energy range.

\ack
This work was supported by Polish Grant from Narodowe Centrum Nauki 2011/01/M/ST9/01891

\end{document}